# Strong polarization mode coupling in microresonators

Sven Ramelow[1,2,*], Alessandro Farsi[1], Stéphane Clemmen[1], Jacob S. Levy[3], Adrea R. Johnson[1],
Yoshitomo Okawachi[1], Michael. R. E. Lamont[1,4], Michal Lipson[3,4], and Alexander L. Gaeta[1,4]

[1]School of Applied and Engineering Physics, Cornell University, Ithaca, New York 14853, USA
[2]Faculty of Physics, University of Vienna, 1090 Vienna, Austria
[3]School of Electrical and Computer Engineering, Cornell University, Ithaca, New York 14853, USA
[4]Kavli Institute at Cornell for Nanoscale Science, Cornell University, Ithaca, New York 14853, USA
*Corresponding author: sven.ramelow@univie.ac.at

We observe strong modal coupling between the $TE_{00}$ and $TM_{00}$ modes in $Si_3N_4$ ring resonators revealed by avoided crossings of the corresponding resonances. Such couplings result in significant shifts of the resonance frequencies over a wide range around the crossing points. This leads to an effective dispersion that is one order of magnitude larger than the intrinsic dispersion and creates broad windows of anomalous dispersion. We also observe the changes to frequency comb spectra generated in $Si_3N_4$ microresonators due polarization mode and higher-order mode crossings and suggest approaches to avoid these effects. Alternatively, such polarization mode-crossings can be used as a novel tool for dispersion engineering in microresonators.

Optical microresonators are important for a wide range of applications, such as parametric frequency combs [1-10], optomechanics [11,12], and in quantum optics as sources for photon-pairs [13-19] or squeezed states [20-21]. The microresonator resonances can in principle be precisely calculated using the dispersion of the resonating modes and the resonator length. However, modal coupling between different types of modes can significantly alter the shape and position of their resonances. Mode splitting occurs for strong coupling [22], and coupling between whole families of modes results in avoided crossings [23-27]. This can lead to dramatic localized changes in the effective dispersion near these crossing points, which in general affects any parametric interaction that relies on precise frequency matching of different resonances. In particular it can play an important role in the formation of parametric frequency combs [24-31]. While mode-crossings can be disruptive for comb generation by inhibiting soliton formation [25] and distorting the comb spectrum [27], they can also be beneficial, allowing for comb formation in resonators with normal group-velocity dispersion (GVD) [8,24] or aiding the generation of dark solitons in normal GVD resonators [29]. In the context of frequency comb generation, only modal interactions between different families of spatial modes have been considered thus far. However, in dielectric waveguides, even when the waveguide is 'single mode', there are typically at least two guided fundamental modes, the fundamental quasi transverse electric ($TE_{00}$) and the fundamental quasi transverse magnetic ($TM_{00}$) mode, which correspond approximately to the polarization of light in the waveguide.

Here, we report on the observation of avoided crossings that result from the strong modal coupling between the $TE_{00}$ and $TM_{00}$ polarization modes in $Si_3N_4$ microring resonators. Similarly, strong polarization mode coupling has been shown to be useful for polarization conversion based on silicon oxinitride technology [31]. Since such a mode interaction can even occur in single-mode waveguides, it is more fundamental than other forms of modal interactions (i.e., between higher-order spatial modes). The physical origin and strength of the modal coupling between the $TE_{00}$ and $TM_{00}$ modes are based on different parameters of the ring resonator such as its radius of curvature, waveguide cross-section, and side-wall angle [32,33]. Microresonators with smaller radii and larger side-wall angles typically will exhibit greater modal coupling.

Our experimental setup for investigating polarization mode coupling is depicted in Fig. 1. We probe the resonators with two different external-cavity diode lasers covering a total tuning range between 1450 nm and 1640 nm. Lensed fibers are used to couple into and out of the bus waveguide with inverted tapers [34] for mode-matching. The polarization of the input and output light is controlled and analyzed with standard fiber-based polarization controllers and a polarization beam splitter, and the output power is monitored with a sensitive photodiode. We use a temperature controller with a Peltier element on the chip holder to stabilize and tune the $Si_3N_4$ microring resonators under investigation. To overcome the limited precision of our tunable lasers, we use an automated stepped scanning and fitting routine supplemented by calibrating each resonance position with a high-precision wavemeter. We find that this method leads to an average precision better than 50 MHz.

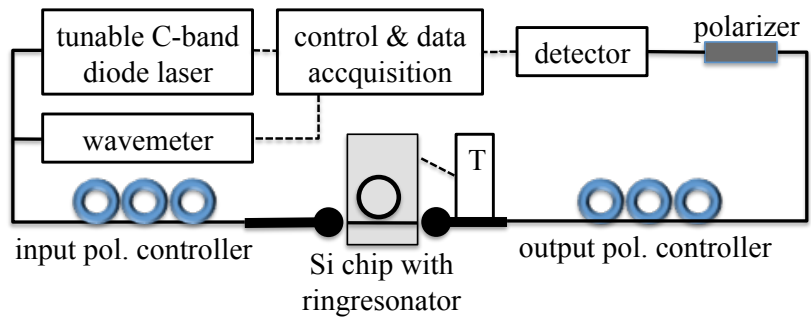

Fig. 1. Schematic of the setup used for observing and characterizing polarization mode coupling in mircoresonators on a temperature stabilized (T) silicon chip.

We first investigate polarization mode coupling in two $Si_3N_4$ microrings with $725\times1100$ $nm^2$ and $725\times900$ $nm^2$ waveguide cross-section and a 100 μm radius. The transmission measurement for $TM_{00}$ input light for the first microring (Fig. 2) yields sharp and deep resonances due to nearly critical coupling between the bus waveguide and the resonator. Near 1595 nm a second sharp resonance, which we associate with the $TE_{00}$ mode, appears on the right side of the main $TM_{00}$ resonance and becomes deeper until both show the same extinction. The main resonance then experiences an adiabatic crossover and the secondary resonance (now on the left side) slowly disappears. We attribute this behavior to an avoided crossing at 1595 nm associated with a strong modal interaction between the $TE_{00}$ and $TM_{00}$ modes. We verify this by a number of different measurements. First, we observe that the adiabatic crossover behavior is inverted when the initial input is $TE_{00}$ polarized. Additionally, we verified that the eigenmodes at the crossing are fully hybridized being a superposition of the $TE_{00}$ and $TM_{00}$ polarization.

To observe the avoided crossing associated with the strong polarization mode coupling, we next precisely measured the resonance wavelengths for both the $TE_{00}$ and $TM_{00}$ polarizations as shown in Fig. 3. For the first microring the measured free spectral ranges (FSR's) at the start of the scan (1510 nm) are 225.0 GHz and 226.3 GHz for the $TM_{00}$ and $TE_{00}$ modes, respectively. This agrees very well with the FSR's calculated from the simulated dispersion based on a finite-element mode-solver [also shown in Fig. 2(b)] and therefore further corroborates the identification of the modes as $TM_{00}$ and $TE_{00}$. As shown in Fig. 3(a) a strong avoided crossing occurs near 1595 nm with the upper branch (blue) changing its mode character continuously from $TE_{00}$ to $TM_{00}$, and vice versa for the second branch. We find a similarly strong avoided crossing for the microring with a $725\times900$ $nm^2$ waveguide cross-section [Fig. 3(d)]. The splittings of the modes at the anti-crossings is 7.2 GHz and 12.5 GHz, respectively. For both resonators this is around 20 times larger than their intrinsic loss rates, which we measure to be 340 MHz ($Q_{int} \approx 600{,}000$) and 700 MHz ($Q_{int} \approx 300{,}000$), respectively, and shows that for both microrings the $TE_{00}$ and $TM_{00}$ modes are strongly coupled.

As shown in Fig. 3(b) and 3(e), near the crossing points, the measured FSR's deviate significantly from their values given by the dispersion and resonator length. This can be interpreted as a mode-coupling induced effective dispersion [8]. By fitting the FSR's for the different branches and taking the derivatives of these fits the values of this effective GVD can be directly determined from the measured data [Fig. 3(c),(f)]. For the first microring we observe relatively high values of the effective GVD reaching around +600 ps/nm/km and -900 ps/nm/km, respectively. Moreover, for the $TE_{00}$-branch (that turns into $TM_{00}$ after the crossing) there is a wide range between 1550 nm and 1610 nm of anomalous GVD. We measure a similar effect on the effective GVD in the second microring [Fig 3(f)] with an anomalous GVD region between 1450 nm and 1495 nm.

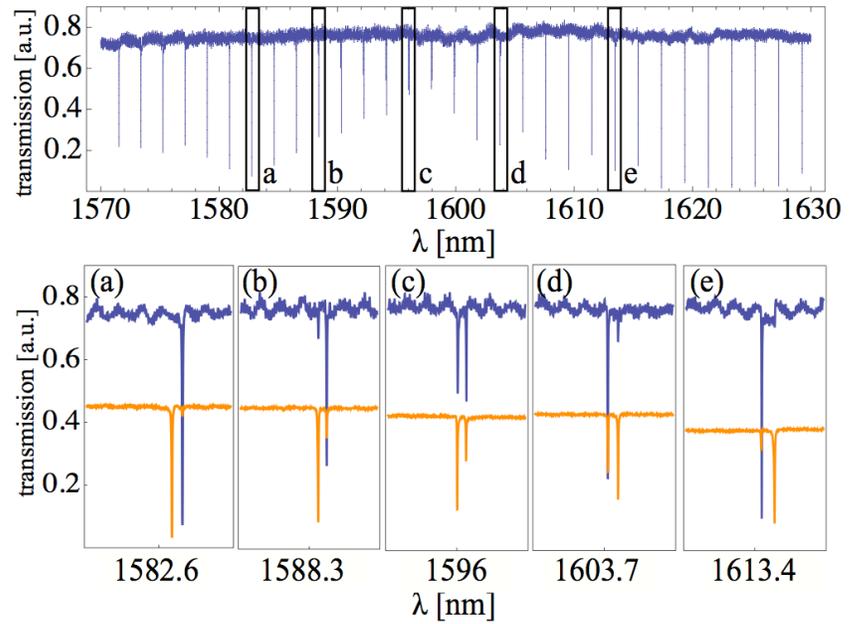

Fig. 2. Optical transmission for $TM_{00}$ polarized light and selected resonances (a) – (e) in detail for $TM_{00}$ (blue) and $TE_{00}$ (orange). The appearance of the $TE_{00}$ resonances for $TM_{00}$ polarized input (and vice versa) and their behavior indicate an avoided crossing near 1595 nm.

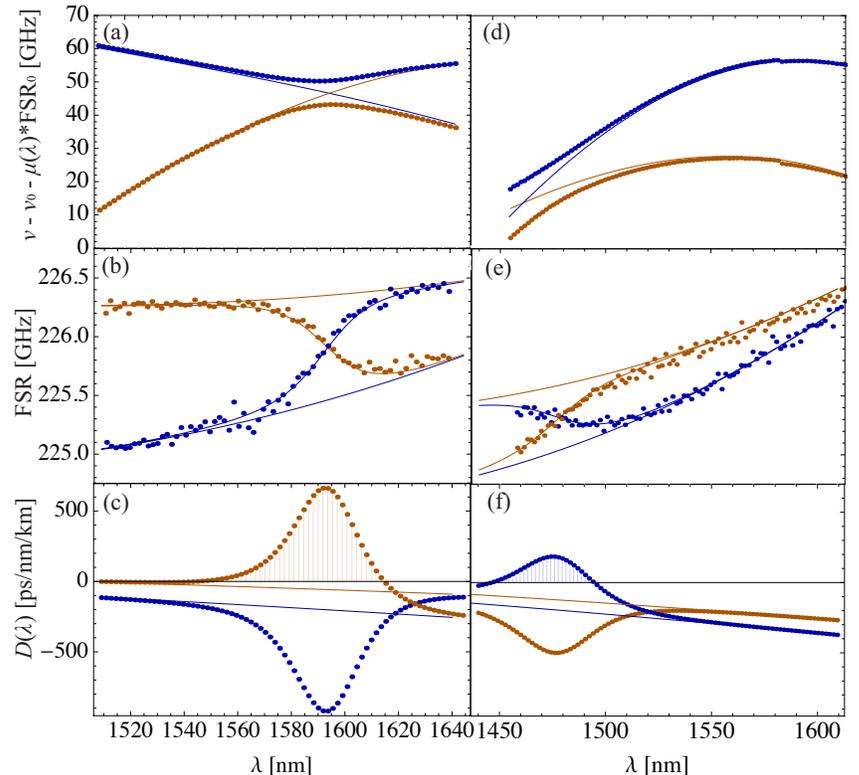

Fig. 3. Polarization avoided crossings for two $Si_3N_4$ microrings with $725\times1100$ $nm^2$ cross-section [(a)-(c)] and $725\times900$ $nm^2$ cross-section [(d)-(f)]. (a),(d) Measured resonance frequencies $v$ of the $TE_{00}$ and $TM_{00}$ modes parameterized by their relative mode number $\mu$, a constant $FSR_0 = 226$ GHz and an offset frequency of $v_0 = c/1509.2$ nm for (a) and $v_0 = c/1456.4$ nm for (d). The solid lines show the theoretical $TE_{00}$ and $TM_{00}$ resonance frequencies based on simulations of the eigenmodes inside the resonator without accounting for the modal interaction. (b),(e) Measured FSRs (dotted), fits to the measured data (solid lines) and the simulation results. (c),(f) Experimentally determined effective GVD (dotted) derived from the fits to the measured data in (b) and (e), respectively, and simulation results yielding the intrinsic GVD (solid lines).

We further investigate the effects of both polarization and higher-order mode crossings on comb generation by directly comparing the comb spectra generated in microring resonators with a measurement of its mode-crossings (Fig. 4). The combs are generated by strongly pumping a single resonance at 1540 nm [4]. We investigate a microring resonator with a waveguide cross-section of 725×1650 $nm^2$ and a 100-µm radius [Fig. 4(a)]. We observe a polarization mode-crossing near 1580 nm with a splitting at the avoided crossing of 2.2 GHz, which greatly exceeds the intrinsic loss rate of the resonator. This polarization mode crossing manifests as a reduction in the mode intensity in the generated comb spectrum. In addition, there exists an anomaly in the FSR at 1550 nm, which can be attributed to a mode crossing with a higher-order spatial mode, and we observe a similar corresponding feature in the comb spectrum. Furthermore, we observe suppressed comb generation when pumping near one of the higher order mode crossings which we attribute to the large change in the effective dispersion. Since this effect can be disadvantageous for applications related to frequency comb generation, it needs to be taken into account in the optimal device design. Moreover, we investigate the effects of mode crossings on comb generation in an 80-GHz FSR microresonator with a waveguide cross-section of 725×1700 $nm^2$ and 1.8-mm length [Fig. 4(b)]. For this microresonator our measurements reveal three polarization mode crossings and a number of higher-order mode crossings that affect the comb spectrum generated from this resonator.

We identify different strategies that could minimize the disruptive effects of mode-crossings (polarization and higher-order mode) on frequency comb generation. In general, mode-crossings are especially disruptive when they appear directly at the pump wavelength or when they line up symmetrically with respect to the pump since they will then simultaneously affect both the signal and idler resonances. Since the position of any mode-crossing depends sensitively on the exact length of the microresonators, slight variations in the design of the resonator lengths can be used to circumvent both scenarios. In addition, if the target application allows it, increasing the FSR will reduce the frequency at which mode-crossings will occur. Another approach for reducing mode-crossing disruptions would be to minimize the modal interactions altogether, which may be achieved by optimizing the microresonator design and fabrication, including larger bend radii or smaller side-wall angles. Furthermore, we have observed experimentally that higher-order mode-crossing are suppressed for the narrower and thus more symmetric cross-section resonators (Fig. 3) as compared to the wider, more asymmetric cross-section resonators (Fig. 4). This indicates that the aspect ratio can be used as an additional parameter for controlling mode-crossing effects.

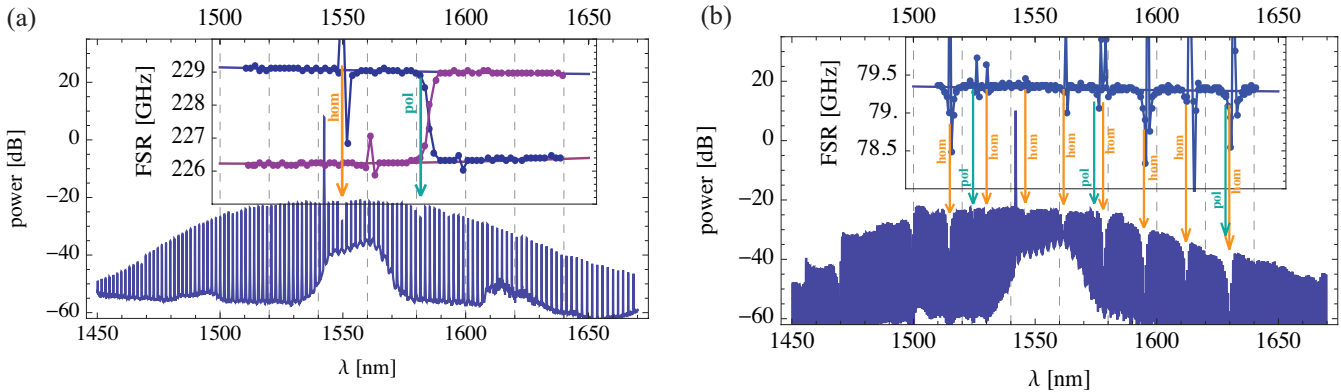

Fig. 4 Generated comb spectra in $Si_3N_4$ microresonators. Insets show the measured FSRs with the (upper) blue branch being for the $TE_{00}$ mode and revealing the mode crossings. (a) Results for a 725×1650 $nm^2$ cross-section, 100-µm radius microring resonator. A polarization mode crossing (pol) occurs near 1580 nm with a corresponding feature (indicated by an arrow) in the comb spectrum. A second anomaly in the FSR's near 1550 nm due to a mode crossing with a higher order spatial mode (hom), also produces a corresponding feature in the spectrum. (b) Results for a 725×1700 $nm^2$ cross-section, 1.8-mm length microresonator. A number of polarization and higher-order mode crossings affect the generated comb spectrum.

In summary, we observe and characterize strong avoided crossings between the $TE_{00}$ and $TM_{00}$ resonance frequencies in $Si_3N_4$ micro-resonators and study their effect on effective GVD and frequency comb generation. We believe there are several interesting applications of strong polarization crossings. First, for certain applications such as non-degenerate frequency conversion involving a strong pump field is useful to suppress parametric processes solely stimulated by the pump, as this would induce noise for the target process. The frequency-mismatch induced by mode crossings can be very strong (we observe an effective GVD as large as 900 ps/nm/km), which can used for suppression of unwanted parametric processes. In addition, we measure a rather smooth and well-defined change in the effective dispersion and observe wide windows of anomalous GVD in otherwise normal GVD microresonators. These anomalous GVD windows could allow the generation of parametric frequency combs in such normal GVD resonators. Moreover, the $TE_{00}$ and $TM_{00}$ modes can simultaneously experience significantly lower loss than the higher-order modes. The exact position of polarization mode-crossings can be controlled by accurately designing the cavity length. Thus, while avoided crossings may be disruptive for certain applications, these polarization mode crossings can be a useful dispersion engineering tool to tailor and optimize microresonator frequency-matching for a wide range of parametric processes underlying frequency comb generation and other microresonator applications.

We acknowledge support from the Defense Advanced Research Projects Agency via the QuASAR program and the Air-Force Office of Scientific Research under grant FA9550-12-1-0377, and Semiconductor Research Corporation. This work was performed in part at the Cornell Nano-Scale Facility, a member of the National Nanotechnology Infrastructure Network, which is supported by the National Science Foundation (grant ECS-0335765). SR is funded by a EU Marie-Curie Fellowship (PIOF-GA-2012- 329851).

**References**
[1]  P. Del'Haye, A. Schliesser, O. Arcizet, T. Wilken, R. Holzwarth, and T. J. Kippenberg, *"Optical frequency comb generation from a monolithic microresonator,"* Nature 450, 1214–1217 (2007).
[2]  A. A. Savchenkov, A. B. Matsko, V. S. Ilchenko, I. Solomatine, D. Seidel, and L. Maleki, *"Tunable optical frequency comb with a crystalline whispering gallery mode resonator,"* Phys. Revl. Lett. **101**, 093902 (2008).
[3]  M. A. Foster, J. S. Levy, O. Kuzucu, K. Saha, M. Lipson, and A.L. Gaeta , *"Silicon-based monolithic optical frequency comb source,"* Opt. Express **19**, 14233 (2011).
[4]  Y. Okawachi, K. Saha, J. S. Levy, Y. H. Wen, M. Lipson, and A. L. Gaeta, *"Octave-spanning frequency comb generation in a silicon nitride chip,"* Opt. Lett., **36**, 3398 (2011).
[5]  W. Liang, A. A. Savchenkov, A. B. Matsko, V. S. Ilchenko, D. Seidel, and L. Maleki, *"Generation of near-infrared frequency combs from a $MgF_2$ whispering gallery mode resonator,"* Opt. Lett. **36**, 2290 (2011).
[6]  S.B. Papp, and S.A. Diddams, *"Spectral and temporal characterization of a fused-quartz-microresonator optical frequency comb,"* Phys. Rev. A **84**, 053833 (2011).
[7]  F. Ferdous, H. Miao, D. E. Leaird, K. Srinivasan, J. Wang, L. Chen, L. T. Varghese, and A. M. Weiner *"Spectral line-by-line pulse shaping of on-chip microresonator frequency combs,"* Nature Photon. **5**, 770 (2011).
[8]  T. Herr, K. Hartinger, J. Riemensberger, C. Y. Wang, E. Gavartin, R. Holzwarth, M. L. Gorodetsky, and T. J. Kippenberg, *"Universal formation dynamics and noise of Kerr-frequency combs in microresonators,"* Nature Photon. **6**, 480 (2012).
[9]  K. Saha, Y. Okawachi, B. Shim, J. S. Levy, R. Salem, A. R. Johnson, M. A. Foster, M. R. E. Lamont, M. Lipson, and A. L. Gaeta, *"Modelocking and femtosecond pulse generation in chip-based frequency combs,"* Opt. Express **21**, 1335 (2013).
[10] T. J. Kippenberg, R. Holzwarth, and S. A. Diddams, *"Microresonator-based optical frequency combs,"* Science **332**, 555 (2011).
[11] T. J. Kippenberg, and K. J. Vahala, *"Cavity Optomechanics: Back-Action at the Mesoscale."* Science 321, 1172–1176 (2008).
[12] M. Aspelmeyer, T. J. Kippenberg, and F. Marquardt, *"Cavity Optomechanics,"* arXiv:1303.0733 (2013).
[13] S. Clemmen, K. Phan Huy, W. Bogaerts, R. G. Baets, Ph. Emplit, and S. Massar, *"Continuous wave photon pair generation in silicon-on-insulator waveguides and ring resonators,"* Opt. Express **17**, 16558 (2009).
[14] S. Clemmen, A. Farsi, J. Levy, L. Helt, M. Liscidini, J. Sipe, M. Lipson, and A. L. Gaeta *"On-chip spectrally-bright photon-pair source from SiN ring micro-cavity,"* in *Frontiers in Optics 2011/Laser Science XXVII*, OSA Technical Digest (Optical Society of America, 2011), paper FThE5.
[15] S. Azzini, D. Grassani, M. J. Strain, M. Sorel, L. G. Helt, J. E. Sipe, M. Liscidini, M. Galli, and D. Bajoni, *"Ultra-low power generation of twin photons in a compact silicon ring resonator,"* Opt. Express **20**, 23100 (2012).
[16] W. C. Jiang, X. Lu, J. Zhang, O. Painter, and Q. Lin, *"A silicon-chip source of bright photon-pair comb"*, arXiv:1210.4455 (2012).
[17] R. Kumar, J. R. Ong, J. Recchio, K. Srinivasan, and S. Mookherjea, *"Spectrally multiplexed and tunable-wavelength photon pairs at 1.55 μm from a silicon coupled-resonator optical waveguide,"* Opt. Lett. **38**, 2969 (2013).
[18] C. Reimer, L. Caspani, M. Clerici, M. Ferrera, M. Kues, M. Peccianti, A. Pasquazi, L. Razzari, B.E. Little, S. T. Chu, D. J. Moss, and R. Morandotti, *"Integrated frequency comb source of heralded single photons,"* Opt. Express **22**, 6535 (2014).
[19] M. Förtsch, J. Fürst, C. Wittmann, D. Strekalov, A. Aiello, M. V. Chekhova, C. Silberhorn, G. Leuchs, and C. Marquardt, *"A versatile source of single photons for quantum information processing,"* Nature Commun. 4, 1818 (2013).
[20] A. H. Safavi-Naeini, S. Gröblacher, J. T. Hill, J. Chan, M. Aspelmeyer, and O. Painter *"Squeezed light from a silicon micromechanical resonator,"* Nature 500, 185–189 (2013).
[21] A. Dutt, K. Luke, S. Manipatruni, A.L. Gaeta, P. Nussenzveig, and M. Lipson, *"On-chip optical squeezing,"* arXiv: 1309.6371, (2013).
[22] T. J. Kippenberg,  S. M. Spillane, and K. J. Vahala, *"Modal coupling in traveling-wave resonators,"* Opt. Lett. **27**, 1669 (2002).
[23] T. Carmon, H. G. L. Schwefel, L. Yang, M. Oxborrow, A. D. Stone, and K. J. Vahala, *"Static envelope patterns in composite resonances generated by level crossing in optical toroidal microcavities,"* Phys. Rev. Lett. **100**, 103905 (2008).
[24] A. A. Savchenkov, A. B. Matsko, W. Liang, V. S. Ilchenko, D. Seidel, and L. Maleki, *"Kerr frequency comb generation in overmoded resonators,"* Opt. Express **20**, 27290 (2012).
[25] T. Herr, V. Brasch, J.D. Jost, I. Mirgorodskiy, G. Lihachev, M.L. Gorodetsky, and T.J. Kippenberg *"Mode spectrum and temporal soliton formation in optical microresonators,"* arXiv:1311.1716 (2013).
[26] Y. Liu, Y. Xuan, X. Xue, P.-H. Wang, A. J. Metcalf, S. Chen, M. Qi, and A. M. Weiner *"Investigation of mode interaction in optical microresonators for Kerr frequency comb generation,"* arXiv:1402.5686 (2014).
[27] I. S. Grudinin, L. Baumgartel, and N. Yu, *"Impact of cavity spectrum on span in microresonator frequency combs,"* Opt. Express **21**, 26929 (2013).
[28] I. S. Grudinin, L. Baumgartel, and N. Yu, *"Frequency comb from a microresonator with engineered spectrum,"* Opt. Express **20**, 6604 (2012).
[29] X. Xue, Y. Xuan, Y. Liu, P.-H. Wang, S. Chen, J. Wang, D. E. Leaird, M. Qi, and A. M. Weiner, *"Mode interaction aided soft excitation of dark solitons in normal dispersion microresonators and offset-frequency tunable Kerr combs,"* arXiv:1404.2865 (2014).
[30] S.-W. Huang, J. F. McMillan, J. Yang, A. Matsko, H. Zhou, M. Yu, D.-L. Kwong, L. Maleki, and C. W. Wong *"Direct generation of 74-fs mode-locking from on-chip normal dispersion frequency combs,"* arXiv:1404.3256 (2014).
[31] A. Melloni, F. Morichetti, and M Martinelli, *"Polarization conversion in ring resonator phase shifters,"* Opt. Lett. 29, 2785–2787 (2004).
[32] W. W. Lui, T. Hirono, K. Yokoyama, and W. Huang, *"Polarization Rotation in Semiconductor Bending  Waveguides: A Coupled-Mode Theory  Formulation,"* J. Lightwave Technol. 16, 929 (1998).
[33] N. Somasiri, and  B. M. A. Rahman, *"Polarization Crosstalk in High Index Contrast Planar Silica Waveguides With Slanted Sidewalls,"* J. Lightwave Technol. 21, 54 (2003).
[34] V. R Almeida, R. R. Panepucci, and M. Lipson, *"Nanotaper for compact mode conversion,"* Opt. Lett. 28, 1302–1304 (2003).